\documentclass[aps,prx,superscriptaddress,reprint]{revtex4-1}
\usepackage{amsmath}
\usepackage{amssymb}
\usepackage{graphicx}
\usepackage{url}
\usepackage{hyperref}
\usepackage{color,xcolor}
\usepackage{epstopdf}
\usepackage{float}
\usepackage{ulem}
\usepackage[percent]{overpic}
\usepackage{siunitx}

\begin{document}

\title{Origin of insulating ferromagnetism in iron oxychalcogenide Ce$_2$O$_2$FeSe$_2$}
\author{Ling-Fang Lin}
\author{Yang Zhang}
\affiliation{Department of Physics and Astronomy, University of Tennessee, Knoxville, Tennessee 37996, USA}
\author{Gonzalo Alvarez}
\affiliation{Computational Sciences \& Engineering Division and Center for Nanophase Materials Sciences, Oak Ridge National Laboratory, Oak Ridge, TN 37831, USA}
\author{Adriana Moreo}
\author{Elbio Dagotto}
\affiliation{Department of Physics and Astronomy, University of Tennessee, Knoxville, Tennessee 37996, USA}
\affiliation{Materials Science and Technology Division, Oak Ridge National Laboratory, Oak Ridge, Tennessee 37831, USA}

\begin{abstract}
An insulating ferromagnetic (FM) phase exists in the quasi-one-dimensional iron oxychalcogenide Ce$_2$O$_2$FeSe$_2$ but its origin is unknown. To understand the FM mechanism, here a systematic investigation of this material is provided, analyzing the competition between ferromagnetic and antiferromagnetic tendencies and the interplay of hoppings, Coulomb interactions, Hund's coupling, and crystal-field splittings. Our intuitive analysis based on second-order perturbation theory shows that large entanglements between doubly-occupied and half-filled orbitals play a key role in stabilizing the FM order in Ce$_2$O$_2$FeSe$_2$. In addition, via many-body computational techniques applied to a multi-orbital Hubbard model, the phase diagram confirms the proposed FM mechanism.
\end{abstract}

\maketitle
\begin{figure}
\centering
\includegraphics[width=0.48\textwidth]{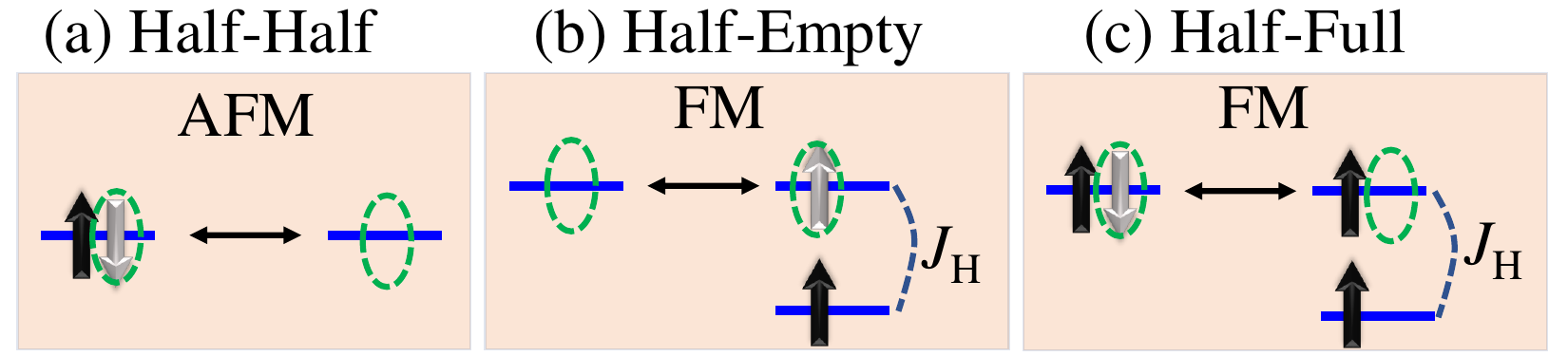}
\caption{Sketch of the basic superexchange cases known (a-c). Blue lines and black arrows represent the orbitals and electrons with spin up or down, respectively. The two-way thin arrows indicated the overlap between inter-site orbitals. Grey thick arrows in the green dashed circles indicate virtual hopping processes. Real materials might host more than one exchange, introducing competition between them~\cite{threemechanism}.}
\label{three_mechanism}
\end{figure}

\textit{Introduction.-}
The understanding of the superexchange phenomenon in transition metal compounds continues 
attracting the attention of the Condensed Matter community \cite{anderson1959new,goodenough1963magnetism,anderson1963theory,lin2021oxygen}. Superexchange theory based on atomic-limit second-order perturbation theory in the hopping amplitudes dominates for many insulators~\cite{anderson1959new,goodenough1963magnetism,anderson1963theory}: as illustrated in Fig.~\ref{three_mechanism}(a), the Heisenberg coupling between two half-filled orbitals is antiferromagnetic (AFM). However, via third-order perturbation theory the case involving half-filled and empty orbitals, coupled with partially filled orbitals via Hund coupling $J_{\rm H}$, Fig.~\ref{three_mechanism}(b), leads to FM order instead. Being third order, the FM coupling magnitude is smaller than the AFM superexchange. Also at third-order, the case involving full-filled and half-filled orbitals, also coupled with partially filled orbitals via $J_{\rm H}$, Fig.~\ref{three_mechanism}(c), should also be FM~\cite{goodenough1963magnetism,weihe1997quantitative}.


Generally, magnetic insulators exhibit multiple types of active interactions, including the kinetic exchange, normally AFM and dominant, and the direct exchange, normally FM and weak~\cite{anderson1959new}. For stronger FM order, mechanisms such as double exchange are invoked, but the induced state is metallic~\cite{Dagotto:Prp,Mayr:Prl}. Thus, the canonical stereotype is that magnetic insulators are primarily antiferromagnets and magnetic metals primarily ferromagnets.

However, robust FM insulators are known experimentally, such as BiMnO$_3$ with orbital order~\cite{dos2002orbital}, double perovskite Sr$_3$OsO$_6$~\cite{wakabayashi2019ferromagnetism}, $90^{\circ}$ bond CrSiTe$_3$~\cite{zhang2019unveiling} and others~\cite{yan2004ferromagnetism,mehta2015long,zhang2021near,taskin2003ising,das2008electronic,feng2020high,mcguire2015coupling,gong2017discovery,huang2020ferromagnetic,steiner1976theoretical}.
Materials with $180^{\circ}$ and $90^{\circ}$ bond cases can be qualitatively understood by the semiempirical Goodenough-Kanamori-Anderson (GKA) rules~\cite{anderson1950antiferromagnetism,anderson1959new,goodenough1955theory,goodenough1958interpretation,kanamori1959superexchange,goodenough1963magnetism,anderson1963theory}. But other cases are more challenging if the cation-anion-cation bond angle deviates substantially from $180^{\circ}$ or $90^{\circ}$, or if the crystal-field symmetry of the magnetic ion is more complicated than in the octahedral coordination. The many FM insulators found experimentally suggest the GKA rules are probably incomplete.

In this publication, we explain the FM insulating state in the quasi-one-dimensional (Q1D) Ce$_2$O$_2$FeSe$_2$ (COFS) material. The new concept we introduce is that, due to geometrical reasons, some of the {\it inter-orbital} electronic hopping amplitudes can become comparable, or even larger, than the intra-orbital hoppings, allowing for FM order to dominate over AFM order. We show explicitly that for COFS, at robust $J_{\rm H}$, remarkably FM order defeats AFM order unveiling a non-canonical mechanism to generate a FM insulator. Our study only requires second-order perturbation theory and thus is directly comparable in magnitude to Anderson's superexchange.

\begin{figure*}
\centering
\includegraphics[width=0.96\textwidth]{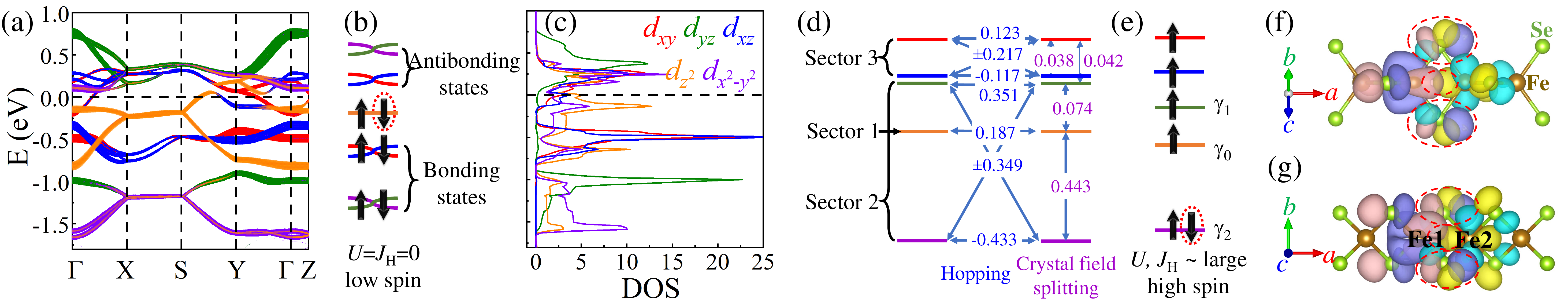}
\caption{(a) DFT iron $3d$ orbital-resolved band structure. (b) Sketches of bonding and antibonding states at small $U$ and $J_{\rm H}$, and their close relation with the band structure in (a). The total population of electrons considered is 6 electrons to fill the energy levels (thick arrows). (c) DOS, from DFT calculations, for the non-magnetic phase. (d) Crude sketches of the crystal-field splitting and dominant NN hopping parameters for the five orbitals. (e) Orbitals and their population at large $U$ and $J_{\rm H}$. The spin down marked with a dashed oval plays a key role in the FM mechanism described in this publication.(f)-(g) Effective Wannier functions (WF) of orbital $\gamma_1$ for Fe1 (in pink and purple) and $\gamma_2$ for Fe2 (in blue and yellow). The robust overlap between these two WFs, related to the amplitude of hoppings $t_{12}$, are indicated by the dashed red ovals.}
\label{analyze_orbital}
\end{figure*}

\textit{Model system.-}
Low dimensional materials and models attract considerable attention ~\cite{kuroki2008unconventional,pandey2021origin,lin2021orbital,lin2019frustrated,zhang2021peierls,zhang2019magnetic,zhang2020iron,zhang2020first,herbrych2020block,pandey2020prediction,patel2019fingerprints,herbrych2019novel}. Experiments showed that Q1D COFS, structurally related to the $Ln$FeAsO family, has a large magnetic moment $\sim 3.14-3.33~\mu_{\rm B}$ on Fe$^{2+}$ ($3d$ $n=6$ electrons) and is ferromagnetically coupled along the dominant chain direction~\cite{mccabe2011new,mccabe2014magnetism}. Each chain is made of distorted edge-sharing FeSe$_4$ tetrahedra and the Fe-Se-Fe bond is $71^{\circ}$, highly deviating from $90^{\circ}$. Experiments showed that the Fe-Se-Ce interactions are much weaker than the Fe-Se-Fe nearest-neighbor (NN) interactions. Thus, COFS is essentialy a 1D chain system~\cite{mccabe2014magnetism}, with weak coupling between chains. The experimental gap $\sim 0.64$ eV shows this material is insulating~\cite{mccabe2011new}. Thus, COFS with robust magnetic order at $176$~K, is an excellent candidate to study deviations of the GKA rules to explain FM insulators.

\textit{DFT Results.-}
From density functional theory (DFT)~\cite{Kresse:Prb99,Blochl:Prb2,Perdew:Prl08}, the band structure of the non-interacting non-magnetic phase of COFS is more dispersive along the chain direction ($\Gamma$-X / S-Y path) than other directions, indicating dominant one-dimensional behavior along the $k_x$ axis [Figs.~\ref{analyze_orbital} (a,b,c)].
Hopping amplitudes and crystal-field splitting energies for the five $3d$ iron orbitals were extracted using the maximally localized Wannier functions~\cite{marzari1997maximally,mostofi2008wannier90} and they are in Fig.~\ref{analyze_orbital} (d). Including the Hubbard $U$ and magnetism, experiments suggest that COFS should be a Mott insulator with no bands crossing the Fermi level.

As shown in Fig.~\ref{analyze_orbital} (d), the orbital $d_{x^2-y^2}$ has the largest NN intra-orbital hopping and the lowest on-site crystal-field energy level. Besides the intra-orbital hoppings, our results indicate that the inter-orbital hoppings -- i.e. non-zero off-diagonal matrix elements -- are also robust.
These important inter-orbital hoppings lead to orbital entanglement. The five orbitals can be naturally divided into three sectors: sector 1 \{$d_{z^2}$\} (primarily isolated), 2 \{$d_{x^2-y^2}$, $d_{yz}$\}, and 3 \{$d_{xz}$, $d_{xy}$\}. From the DFT hopping amplitudes and crystal fields, the total bandwidth for each sector
is, roughly, $\sim$ 1, 2.3, and 1 eV, respectively, with sector 2 having the largest bandwidth~\cite{comments_hoppings}. Note that Fig.~\ref{analyze_orbital} (d) is only a sketch and should not be confused with the bands of Fig.~\ref{analyze_orbital} (a).

If only the kinetic term and crystal-field splitting were included, i.e. at $U = 0$ and $J_{\rm H} = 0$, sectors 2 and 3 will form entangled bonding and antibonding states with band gaps between them [Fig.~\ref{analyze_orbital} (b)]. The 6 electrons would fill energy levels with 3 spins up and 3 spins down forming a non-magnetic state, with electrons distributed in each sector as \{2, 2, 2\}, respectively. However, at large $U$ and $J_{\rm H}$ [Fig.~\ref{analyze_orbital} (e)] orbitals are decoupled and localized forming a Mott phase, and the 6 electrons fill energy levels with 5 spins up and 1 spin down leading to a high-spin state. Electrons are distributed in each sector as \{1, 3, 2\}, respectively. The extra electron (pointing down) will be in orbital $d_{x^2-y^2}$, due to its lowest crystal-field energy.
Because COFS is an insulator with Fe$^{2+}$ in a high-spin state~\cite{mccabe2011new,mccabe2014magnetism} from this analysis the most relevant three orbitals are \{$d_{z^2}$, $d_{yz}$, $d_{x^2-y^2}$\} from sectors 1 and 2 and are, thus, the basis in our Hamiltonian analysis below. Orbitals \{$d_{z^2}$, $d_{yz}$, $d_{x^2-y^2}$\} are labeled \{$\gamma_0$, $\gamma_1$, $\gamma_2$\} for simplicity.

Figures~\ref{analyze_orbital} (f) and (g) provide an intuitive perspective of the effective orbitals and the overlaps between orbital $\gamma_1$ for Fe1 and $\gamma_2$ for Fe2. These effective orbitals are a combination of the Fe $d$ and the Se $p$ orbitals. The dominant overlaps are mainly contributed by
the Se $p$-orbitals, i.e. Se as the Fe-Fe bridge is crucial in the hybridization. A direct overlap between $d$-orbitals is also observed, due to the short distance ($\sim 2.84$ \AA) between NN irons. The entanglements between orbitals, compatible with the large hopping $t_{12}$, plays a key role in COFS, as discussed below.

\textit{Hubbard Model.-}
To understand the magnetism of COFS, a three-orbital Hubbard model for the Fe chain was constructed \cite{Footnote}, including tight-binding kinetic energy and on-site Coulomb
interaction energy terms $H = H_k + H_{int}$. The kinetic portion is
\begin{eqnarray}
H_k = \sum_{\substack{i\sigma \gamma\gamma'}}t_{\gamma\gamma'}
(c^{\dagger}_{i\sigma\gamma}c^{\phantom\dagger}_{{i+1}\sigma\gamma'}+H.c.)+ \sum_{i\sigma\gamma} \Delta_{\gamma} n_{i\sigma\gamma},
\end{eqnarray}
where the first term represents the electron hopping from orbital $\gamma$ at site $i$ to orbital $\gamma'$ at the NN site $i+1$. For simplicity, only the most important NN hopping amplitudes (eV units) are included~\cite{Supp},
\begin{equation}
\begin{split}
t_{\gamma\gamma'} =
\begin{bmatrix}
          0.187	    &  -0.054	   &       0.020	   	       \\
          0.054	    &   0.351	   &      -0.349	   	       \\
          0.020	    &   0.349	   &      -0.433	
\end{bmatrix}.\\
\end{split}
\end{equation}
$\Delta_{\gamma}$ is the crystal-field splitting of orbital $\gamma$, i.e. $\Delta_{0} = -0.277$, $\Delta_{1} = -0.203$, $\Delta_{2} = -0.720$ eV, respectively. The total kinetic energy bandwidth $W$ is 2.085 eV.

The model's electronic interaction -- intraorbital Hubbard repulsion, interorbital repulsion at different orbitals, Hund's coupling, and pair hopping terms -- is:
\begin{eqnarray}
H_{int}= U\sum_{i\gamma}n_{i\uparrow \gamma} n_{i\downarrow \gamma} +(U'-\frac{J_{\rm H}}{2})\sum_{\substack{i\\\gamma < \gamma'}} n_{i \gamma} n_{i\gamma'} \nonumber \\
-2J_{\rm H}  \sum_{\substack{i\\\gamma < \gamma'}} {{\bf S}_{i,\gamma}}\cdot{{\bf S}_{i,\gamma'}}+J_{\rm H}  \sum_{\substack{i\\\gamma < \gamma'}} (P^{\dagger}_{i\gamma} P_{i\gamma'}+H.c.),
\end{eqnarray}
where $U'=U-2J_{\rm H}$ is used and $P_{i\gamma}$=$c_{i \downarrow \gamma} c_{i \uparrow \gamma}$.

\begin{figure}
\centering
\includegraphics[width=0.45\textwidth]{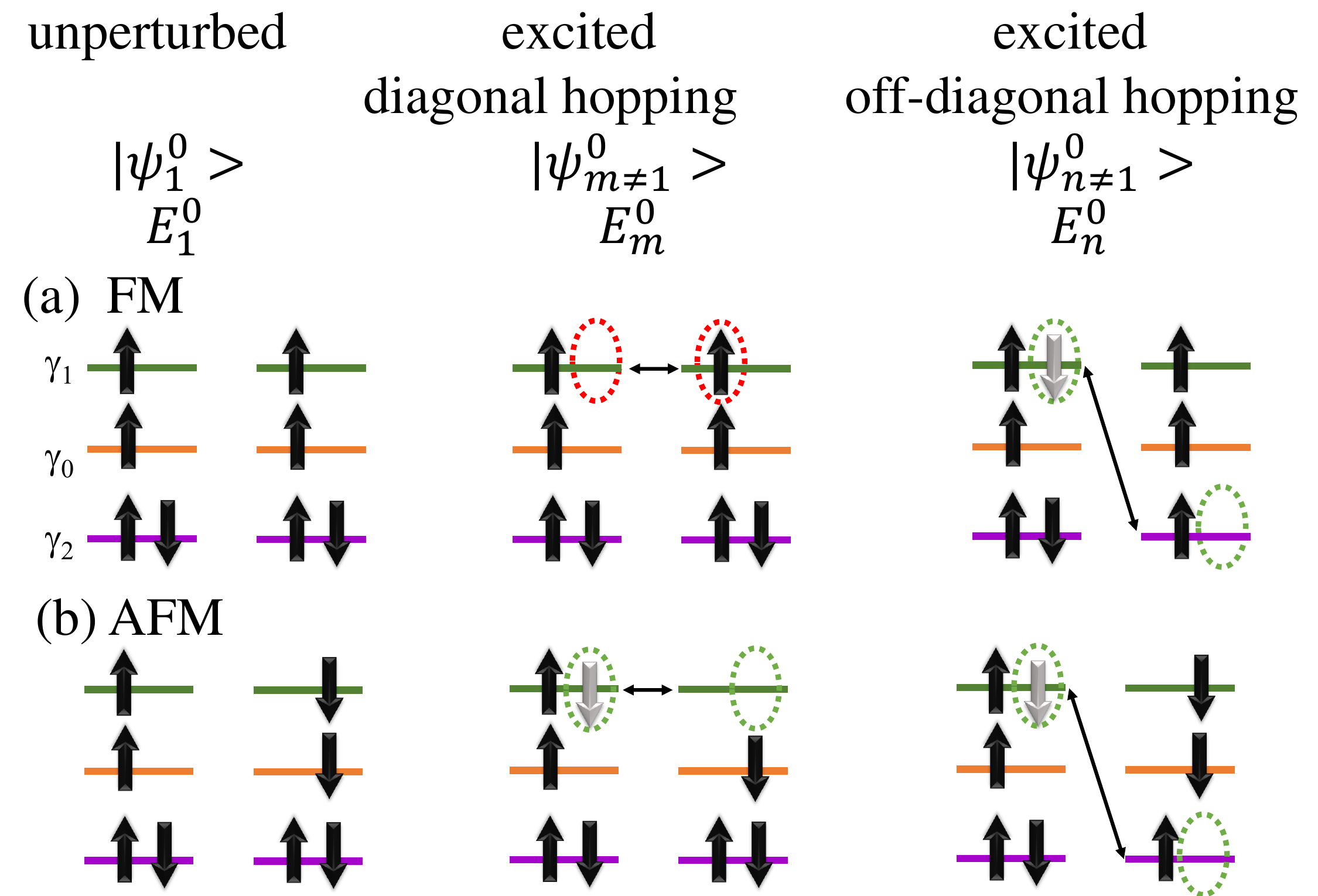}
\caption{Sketches of unperturbed and excited states for diagonal and off-diagonal hoppings. Both FM and AFM cases are considered. For the FM case, the diagonal hopping is forbidden (red ovals) due to Pauli exclusion principle.}
\label{perturbation}
\end{figure}

\textit{Second-order perturbation theory.-} Consider the limit where the hoppings $t_{\gamma \gamma'}{\ll} U, J_{\rm H}$ are the perturbation. A sketch is in Fig. \ref{perturbation}, considering 2 sites and 3 orbitals with population \{1, 1, 2\} and focusing on the $U$, $U'$, $J_{\rm H}$ and $\Delta_{\gamma}$ terms (the pair-hopping is widely considered to play a secondary role in calculations of this kind).
For nondegenerate states, second-order perturbation
theory to the ground state (state 1) always lowers the energy by:
\begin{equation}
\Delta E_1=-\sum_{\substack{m\neq 1}}\frac{|\langle \psi_m^0|H'|\psi_1^0\rangle |^2}{E_m^0-E_1^0}.
\end{equation}

For both the FM and AFM unperturbed states, the ground state atomic energy is
\begin{equation}
E_1^0=2U+10U'-6J_{\rm H}+2{\Delta}_0+2{\Delta}_1+4{\Delta}_2.
\end{equation}

(a) For the FM case, the intra-orbital hopping is forbidden due to the Pauli principle. As for the excited state induced by the off-diagonal hopping, the energy is
\begin{equation}
E_{exc}^0=2U+11U'-7J_{\rm H}+2{\Delta}_0+3{\Delta}_1+3{\Delta}_2.
\end{equation}

By using this second-order perturbation theory, the DFT-deduced hoppings and crystal fields, and
the widely employed ratio $J_{\rm H}/U = 1/4$ for iron superconductors,
the total energy gain of the FM configuration due to $t_{12}$ is
\begin{equation}
\begin{split}
\Delta E_{\rm FM}&=\frac{|\langle \psi_{exc}^0|H'|\psi_1^0\rangle |^2}{E_1^0-E_{exc}^0}\\
&=-\frac{|t_{12}|^2}{U-3J_{\rm H}+{\Delta}_1-{\Delta}_2}.
\end{split}
\end{equation}

(b) For the AFM case, the energy of the diagonal intra-orbital hopping excited state is
\begin{equation}
E_{exc11}^0=3U+10U'-5J_{\rm H}+2{\Delta}_0+2{\Delta}_1+4{\Delta}_2.
\end{equation}

Thus, the energy gain for the AFM configuration due to the intra-orbital hopping $t_{11}$ (or $t_{00}$) is
\begin{equation}
\Delta E_{\rm AFM-11}=\frac{|\langle \psi_{exc}^0|H'|\psi_1^0\rangle |^2}{E_1^0-E_{exc11}^0}=-\frac{|t_{11}|^2}{U+J_{\rm H}}.
\end{equation}

For the same AFM case, the energy of the off-diagonal hopping $t_{12}$ excited state is
\begin{equation}
E_{exc12}^0=2U+11U'-5J_{\rm H}+2{\Delta}_0+3{\Delta}_1+3{\Delta}_2.
\end{equation}

The energy gain for the AFM configuration due to the off-diagonal hopping $t_{12}$ is
\begin{equation}
\begin{split}
\Delta E_{\rm AFM-12}&=\frac{|\langle \psi_n^0|H'|\psi_1^0\rangle |^2}{E_1^0-E_n^0}\\
&=-\frac{|t_{12}|^2}{U-J_{\rm H}+{\Delta}_1-{\Delta}_2}.
\end{split}
\end{equation}

The total energy gained in the AFM state
from the $t_{00}$, $t_{11}$ and $t_{12}$ terms, using the same DFT and Hund parameters as in the FM case, is
\begin{equation}
\begin{split}
\Delta E_{\rm AFM}&=-\frac{|t_{00}|^2}{U+J_{\rm H}}-\frac{|t_{11}|^2}{U+J_{\rm H}}-\frac{|t_{12}|^2}{U-J_{\rm H}+{\Delta}_1-{\Delta}_2}.
\end{split}
\end{equation}

It can be shown that for $U\gtrsim 2.17$ eV, i.e. $U/W\gtrsim 1.04$, $|\Delta E_{FM}|>|\Delta E_{AFM}|$. The large off-diagonal $t_{12}$ hopping plays the key role on the dominance of ferromagnetism 
over antiferromagnetism. Note
the important role of $J_{\rm H}$ as well: repeating the calculation for $J_{\rm H}=0$ the result is reversed and the AFM state wins.
Varying
$J_{\rm H}$, a transition AFM-FM occurs. Specifically, for $U = 10$~eV, the critical ratio is
$J_{\rm H}/U \sim 0.2$ as in the density matrix renormalization group (DMRG) results next.

\begin{figure}
\centering
\includegraphics[width=0.43\textwidth]{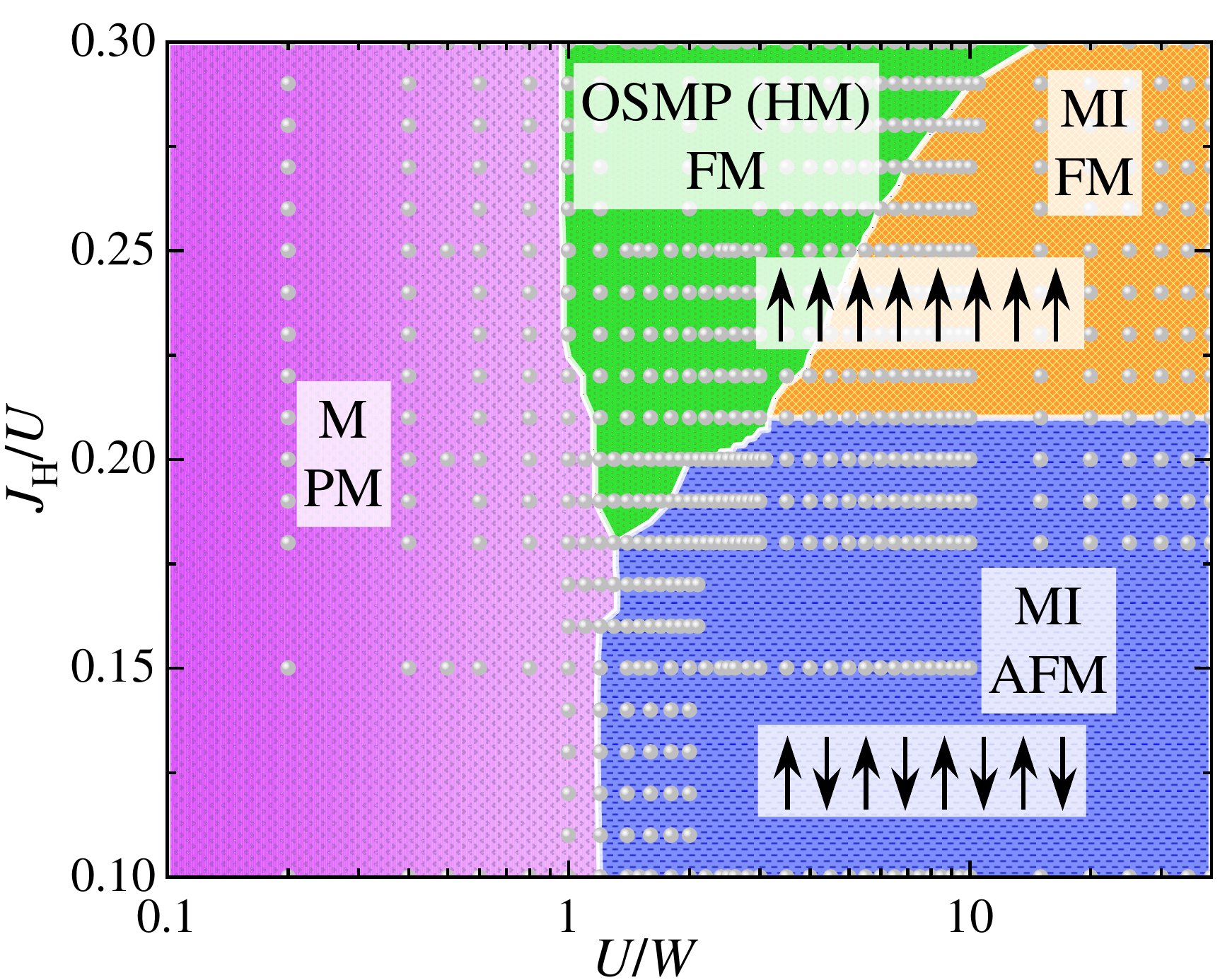}
\caption{DMRG phase diagram of the three-orbital Hubbard model varying $U/W$ and $J_{\rm H}/U$, using a $L=16$ chain. Different phases are indicated, with the conventions metal (M), Hund metal (HM), orbital-selective Mott phase (OSMP), Mott insulator (MI), paramagnetic (PM), antiferromagnetic (AFM), and ferromagnetic (FM) phases. 
Small circles indicate specific
values of data points that were investigated with DMRG. The phase boundaries should only be considered crude approximations because a mixture of competing states as well as incommensurate phases were detected near those boundaries. However, the existence of the four phases shown was clearly established, even if the boundaries are only crude estimations.}
\label{phase}
\end{figure}

\textit{DMRG \& Phase diagram.-}
The magnetic phase diagram (Fig.~\ref{phase}) was calculated varying $U/W$ and $J_{\rm H}/U$ using the DMRG++ code~\cite{white1992density,white1993density,schollwock2005density,hallberg2006new,alvarez2009density}. We found four dominant regimes in our calculations: (1) M-PM, (2) MI-AFM, (3) OSMP-FM, and (4) MI-FM.  At $U/W \lesssim 1$, the system is metallic and non-magnetic (M-PM), as expected. As $U/W$ increases (in particular as $U/W \gtrsim 1$), the system becomes a Mott insulator with AFM spin ordering up-down at $J_{\rm H}/U \lesssim 0.21$ (MI-AFM). Increasing $J_{\rm H}/U$, the system now enters a FM region.  Interestingly, at intermediate $U/W$ and $0.21 \lesssim J_{\rm H}/U \lesssim 0.30$, due to the strong competition between $J_{\rm H}/U$ and $U/W$ the system is in  an exotic OSMP-FM state with the selective localization of electrons on one orbital while
other orbitals remain metallic. This state was much studied recently~\cite{yu2020orbital,patel2019fingerprints,rincon2014exotic}
and will not be discussed here further.
At both large $U/W$ and $J_{\rm H}/U$, the system develops a gap and becomes insulating,
defining the MI-FM state of our focus. From experiments~\cite{mccabe2011new,mccabe2014magnetism}, COFS should be located at the MI-FM phase. Here, the large $U$ renders all orbitals localized and in a high-spin state leading to the unusual FM configuration explained before via second-order perturbation theory.

\begin{figure}
\centering
\includegraphics[width=0.43\textwidth]{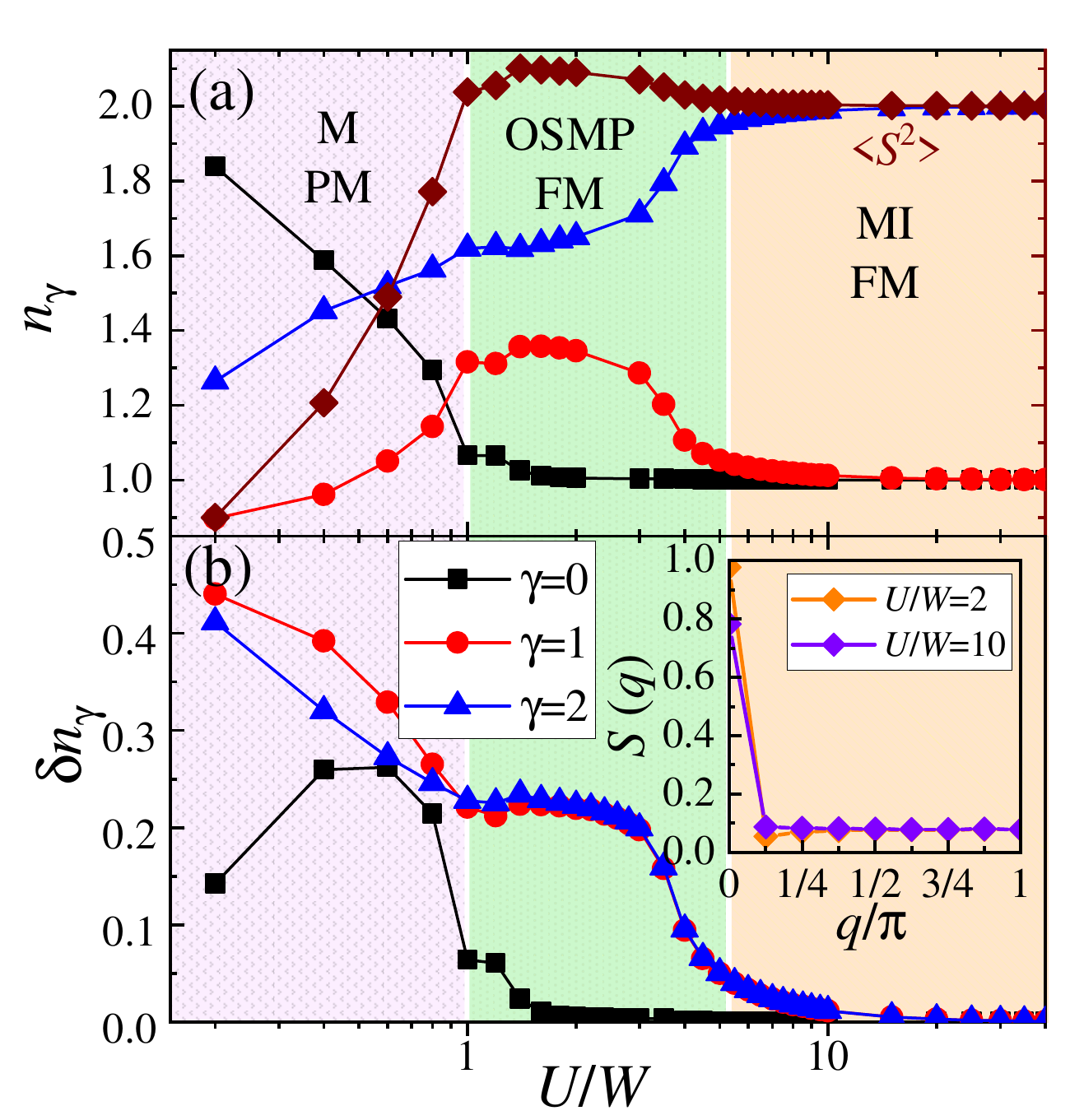}
\caption{(a) Orbital-resolved occupation number $n_{\gamma}$, mean value of the total spin squared $\langle {\bf{S}}^2\rangle $ (in maroon color) and (b) charge fluctuations vs $U/W$ at $J_{\rm H}/U = 1/4$. Inset: spin structure factor for $U/W = 2$ and 10. Similarly as in Fig.~\ref{phase}, at the boundaries phase competition renders some results slightly inaccurate. For example, the first two black points in (a) from the left in the OSMP regime are not exactly 1. Yet because their state is FM, we believe they are part of the OSMP FM rather than a new phase.}
\label{n_u}
\end{figure}

For the prototypical value $J_{\rm H}/U = 1/4$, the electronic occupancy and local moment are in Fig.~\ref{n_u} (a). In the small-$U$ metallic PM phase, the $n_{\gamma}$ values of all three orbitals evolve smoothly from the noninteracting limit with increasing $U$: the extra electron gradually transfers from sector 1 to sector 2 until a critical $U \sim W$. When $U/W \gtrsim 4$ and arriving to the MI-FM phase, the extra electron has totally transferred to ${\gamma}_2$, leading to $n_{2} = 2$, while $n_{0} = n_{1} = 1$. In the MI-FM region, the fluctuations of all three orbitals are suppressed and the total spin squared, Fig.~\ref{n_u} (a), saturates to 2, i.e. spin 1, the maximum number our study can generate (4 electrons in 3 orbitals per site). In this sense, our results agree with powder neutron scattering that also reported a spin close to the maximum possible for 
Fe$^{2+}$~\cite{moment}.

The spin structure factors $S$($q$) for $U/W = 2$ and 10 are in the inset of Fig.~\ref{n_u} (b).
$S$($q$) displays a sharp peak at $q = 0$ for both $U/W = 2$ and 10, indicating FM order within the OSMP and MI regions, the latter as in experiments~\cite{mccabe2011new,mccabe2014magnetism}. The corresponding single-particle spectra and DOS for $J_{\rm H}/U = 1/4$ and $U/W = 10$ are in the SM~\cite{Supp}. Orbitals ${\gamma}_0$ and ${\gamma}_1$ are half-filled with a gap, while orbital ${\gamma}_2$ is full-filled. Then, the final gap becomes $\sim 2.5$ eV.

\textit{Conclusions. -}
The iron oxychalcogenide COFS with $n = 6$ electrons per Fe was studied by DFT, by DMRG applied to a three-orbital model, and by second-order perturbation theory to gain intuitive insight. We showed that in COFS inter-orbital electronic hopping amplitudes can become comparable or larger than the intra-orbital hoppings due to geometrical reasons, stabilizing an insulating FM state.
The proposed mechanism was confirmed via DMRG. Our rich phase diagram suggests that COFS is at strong $U/W$, as in experiments~\cite{mccabe2011new,mccabe2014magnetism}.
The proposed mechanism requires $J_{\rm H}$ to be
robust as well, as in iron supercondcutors. We predict that the similar chain system Na$_2$Fe$X_2$~\cite{stuble2018na7} should also exhibit FM coupling along the chain direction.

\textit{Acknowledgments. -}
The work of L.-F.L., Y.Z., A.M. and E.D. was supported by the U.S. Department of Energy (DOE), Office of Science, Basic Energy Sciences (BES), Materials Sciences and Engineering Division. G.A. was partially supported by the scientific Discovery through Advanced Computing (SciDAC) program funded by U.S. DOE, Office of Science, Advanced Scientific Computing Research and BES, Division of Materials Sciences and Engineering. The calculations were carried out at the University of Tennessee Advanced Computational Facility (ACF).

\bibliographystyle{apsrev4-1}
\bibliography{ref3}

\begin{thebibliography}{58}%
\makeatletter
\providecommand \@ifxundefined [1]{%
 \@ifx{#1\undefined}
}%
\providecommand \@ifnum [1]{%
 \ifnum #1\expandafter \@firstoftwo
 \else \expandafter \@secondoftwo
 \fi
}%
\providecommand \@ifx [1]{%
 \ifx #1\expandafter \@firstoftwo
 \else \expandafter \@secondoftwo
 \fi
}%
\providecommand \natexlab [1]{#1}%
\providecommand \enquote  [1]{``#1''}%
\providecommand \bibnamefont  [1]{#1}%
\providecommand \bibfnamefont [1]{#1}%
\providecommand \citenamefont [1]{#1}%
\providecommand \href@noop [0]{\@secondoftwo}%
\providecommand \href [0]{\begingroup \@sanitize@url \@href}%
\providecommand \@href[1]{\@@startlink{#1}\@@href}%
\providecommand \@@href[1]{\endgroup#1\@@endlink}%
\providecommand \@sanitize@url [0]{\catcode `\\12\catcode `\$12\catcode
  `\&12\catcode `\#12\catcode `\^12\catcode `\_12\catcode `\%12\relax}%
\providecommand \@@startlink[1]{}%
\providecommand \@@endlink[0]{}%
\providecommand \url  [0]{\begingroup\@sanitize@url \@url }%
\providecommand \@url [1]{\endgroup\@href {#1}{\urlprefix }}%
\providecommand \urlprefix  [0]{URL }%
\providecommand \Eprint [0]{\href }%
\providecommand \doibase [0]{http://dx.doi.org/}%
\providecommand \selectlanguage [0]{\@gobble}%
\providecommand \bibinfo  [0]{\@secondoftwo}%
\providecommand \bibfield  [0]{\@secondoftwo}%
\providecommand \translation [1]{[#1]}%
\providecommand \BibitemOpen [0]{}%
\providecommand \bibitemStop [0]{}%
\providecommand \bibitemNoStop [0]{.\EOS\space}%
\providecommand \EOS [0]{\spacefactor3000\relax}%
\providecommand \BibitemShut  [1]{\csname bibitem#1\endcsname}%
\let\auto@bib@innerbib\@empty
\bibitem [{\citenamefont {Anderson}(1959)}]{anderson1959new}%
  \BibitemOpen
  \bibfield  {author} {\bibinfo {author} {\bibfnamefont {P.~W.}\ \bibnamefont
  {Anderson}},\ }\href {https://doi.org/10.1103/PhysRev.115.2} {\bibfield
  {journal} {\bibinfo  {journal} {Phys. Rev.}\ }\textbf {\bibinfo {volume}
  {115}},\ \bibinfo {pages} {2} (\bibinfo {year} {1959})}\BibitemShut {NoStop}%
\bibitem [{\citenamefont {Goodenough}(1963)}]{goodenough1963magnetism}%
  \BibitemOpen
  \bibfield  {author} {\bibinfo {author} {\bibfnamefont {J.~B.}\ \bibnamefont
  {Goodenough}},\ }\href@noop {} {\emph {\bibinfo {title} {Magnetism and the
  Chemical Bond}}},\ Vol.~\bibinfo {volume} {1}\ (\bibinfo  {publisher}
  {Interscience publishers (Wiley), New York},\ \bibinfo {year}
  {1963})\BibitemShut {NoStop}%
\bibitem [{\citenamefont {Anderson}(1963)}]{anderson1963theory}%
  \BibitemOpen
  \bibfield  {author} {\bibinfo {author} {\bibfnamefont {P.~W.}\ \bibnamefont
  {Anderson}},\ }in\ \href {https://doi.org/10.1016/S0081-1947(08)60260-X}
  {\emph {\bibinfo {booktitle} {Solid state physics}}},\ Vol.~\bibinfo {volume}
  {14}\ (\bibinfo  {publisher} {Elsevier},\ \bibinfo {year} {1963})\ pp.\
  \bibinfo {pages} {99--214}\BibitemShut {NoStop}%
\bibitem [{\citenamefont {Lin}\ \emph {et~al.}(2021{\natexlab{a}})\citenamefont
  {Lin}, \citenamefont {Kaushal}, \citenamefont {Sen}, \citenamefont
  {Christianson}, \citenamefont {Moreo},\ and\ \citenamefont
  {Dagotto}}]{lin2021oxygen}%
  \BibitemOpen
  \bibfield  {author} {\bibinfo {author} {\bibfnamefont {L.-F.}\ \bibnamefont
  {Lin}}, \bibinfo {author} {\bibfnamefont {N.}~\bibnamefont {Kaushal}},
  \bibinfo {author} {\bibfnamefont {C.}~\bibnamefont {Sen}}, \bibinfo {author}
  {\bibfnamefont {A.~D.}\ \bibnamefont {Christianson}}, \bibinfo {author}
  {\bibfnamefont {A.}~\bibnamefont {Moreo}}, \ and\ \bibinfo {author}
  {\bibfnamefont {E.}~\bibnamefont {Dagotto}},\ }\href
  {https://doi.org/10.1103/PhysRevB.103.184414} {\bibfield  {journal} {\bibinfo
   {journal} {Phys. Rev. B}\ }\textbf {\bibinfo {volume} {103}},\ \bibinfo
  {pages} {184414} (\bibinfo {year} {2021}{\natexlab{a}})}\BibitemShut
  {NoStop}%
\bibitem [{\citenamefont {Weihe}\ and\ \citenamefont
  {G{\"u}del}(1997)}]{weihe1997quantitative}%
  \BibitemOpen
  \bibfield  {author} {\bibinfo {author} {\bibfnamefont {H.}~\bibnamefont
  {Weihe}}\ and\ \bibinfo {author} {\bibfnamefont {H.~U.}\ \bibnamefont
  {G{\"u}del}},\ }\href {https://doi.org/10.1021/ic961502+} {\bibfield
  {journal} {\bibinfo  {journal} {Inorg. Chem.}\ }\textbf {\bibinfo {volume}
  {36}},\ \bibinfo {pages} {3632} (\bibinfo {year} {1997})}\BibitemShut
  {NoStop}%
\bibitem [{\citenamefont {Dagotto}\ \emph {et~al.}(2001)\citenamefont
  {Dagotto}, \citenamefont {Hotta},\ and\ \citenamefont {Moreo}}]{Dagotto:Prp}%
  \BibitemOpen
  \bibfield  {author} {\bibinfo {author} {\bibfnamefont {E.}~\bibnamefont
  {Dagotto}}, \bibinfo {author} {\bibfnamefont {T.}~\bibnamefont {Hotta}}, \
  and\ \bibinfo {author} {\bibfnamefont {A.}~\bibnamefont {Moreo}},\ }\href
  {https://doi.org/10.1016/S0370-1573(00)00121-6} {\bibfield  {journal}
  {\bibinfo  {journal} {Phys. Rep.}\ }\textbf {\bibinfo {volume} {344}},\
  \bibinfo {pages} {1} (\bibinfo {year} {2001})}\BibitemShut {NoStop}%
\bibitem [{\citenamefont {Mayr}\ \emph {et~al.}(2001)\citenamefont {Mayr},
  \citenamefont {Moreo}, \citenamefont {Verg\'{e}s}, \citenamefont {Arispe},
  \citenamefont {Feiguin},\ and\ \citenamefont {Dagotto}}]{Mayr:Prl}%
  \BibitemOpen
  \bibfield  {author} {\bibinfo {author} {\bibfnamefont {M.}~\bibnamefont
  {Mayr}}, \bibinfo {author} {\bibfnamefont {A.}~\bibnamefont {Moreo}},
  \bibinfo {author} {\bibfnamefont {J.~A.}\ \bibnamefont {Verg\'{e}s}},
  \bibinfo {author} {\bibfnamefont {J.}~\bibnamefont {Arispe}}, \bibinfo
  {author} {\bibfnamefont {A.}~\bibnamefont {Feiguin}}, \ and\ \bibinfo
  {author} {\bibfnamefont {E.}~\bibnamefont {Dagotto}},\ }\href@noop {}
  {\bibfield  {journal} {\bibinfo  {journal} {Phys. Rev. Lett.}\ }\textbf
  {\bibinfo {volume} {86}},\ \bibinfo {pages} {135} (\bibinfo {year}
  {2001})}\BibitemShut {NoStop}%
\bibitem [{\citenamefont {Moreira~dos Santos}\ \emph
  {et~al.}(2002)\citenamefont {Moreira~dos Santos}, \citenamefont {Cheetham},
  \citenamefont {Atou}, \citenamefont {Syono}, \citenamefont {Yamaguchi},
  \citenamefont {Ohoyama}, \citenamefont {Chiba},\ and\ \citenamefont
  {Rao}}]{dos2002orbital}%
  \BibitemOpen
  \bibfield  {author} {\bibinfo {author} {\bibfnamefont {A.}~\bibnamefont
  {Moreira~dos Santos}}, \bibinfo {author} {\bibfnamefont {A.~K.}\ \bibnamefont
  {Cheetham}}, \bibinfo {author} {\bibfnamefont {T.}~\bibnamefont {Atou}},
  \bibinfo {author} {\bibfnamefont {Y.}~\bibnamefont {Syono}}, \bibinfo
  {author} {\bibfnamefont {Y.}~\bibnamefont {Yamaguchi}}, \bibinfo {author}
  {\bibfnamefont {K.}~\bibnamefont {Ohoyama}}, \bibinfo {author} {\bibfnamefont
  {H.}~\bibnamefont {Chiba}}, \ and\ \bibinfo {author} {\bibfnamefont
  {C.~N.~R.}\ \bibnamefont {Rao}},\ }\href {\doibase
  10.1103/PhysRevB.66.064425} {\bibfield  {journal} {\bibinfo  {journal} {Phys.
  Rev. B}\ }\textbf {\bibinfo {volume} {66}},\ \bibinfo {pages} {064425}
  (\bibinfo {year} {2002})}\BibitemShut {NoStop}%
\bibitem [{\citenamefont {Wakabayashi}\ \emph {et~al.}(2019)\citenamefont
  {Wakabayashi}, \citenamefont {Krockenberger}, \citenamefont {Tsujimoto},
  \citenamefont {Boykin}, \citenamefont {Tsuneyuki}, \citenamefont {Taniyasu},\
  and\ \citenamefont {Yamamoto}}]{wakabayashi2019ferromagnetism}%
  \BibitemOpen
  \bibfield  {author} {\bibinfo {author} {\bibfnamefont {Y.~K.}\ \bibnamefont
  {Wakabayashi}}, \bibinfo {author} {\bibfnamefont {Y.}~\bibnamefont
  {Krockenberger}}, \bibinfo {author} {\bibfnamefont {N.}~\bibnamefont
  {Tsujimoto}}, \bibinfo {author} {\bibfnamefont {T.}~\bibnamefont {Boykin}},
  \bibinfo {author} {\bibfnamefont {S.}~\bibnamefont {Tsuneyuki}}, \bibinfo
  {author} {\bibfnamefont {Y.}~\bibnamefont {Taniyasu}}, \ and\ \bibinfo
  {author} {\bibfnamefont {H.}~\bibnamefont {Yamamoto}},\ }\href
  {https://doi.org/10.1038/s41467-019-08440-6} {\bibfield  {journal} {\bibinfo
  {journal} {Nat. Commun.}\ }\textbf {\bibinfo {volume} {10}},\ \bibinfo
  {pages} {1} (\bibinfo {year} {2019})}\BibitemShut {NoStop}%
\bibitem [{\citenamefont {Zhang}\ \emph
  {et~al.}(2019{\natexlab{a}})\citenamefont {Zhang}, \citenamefont {Cai},
  \citenamefont {Xia}, \citenamefont {Liang}, \citenamefont {Huang},
  \citenamefont {Wang}, \citenamefont {Yang}, \citenamefont {Yuan},
  \citenamefont {Chen}, \citenamefont {Zhang} \emph
  {et~al.}}]{zhang2019unveiling}%
  \BibitemOpen
  \bibfield  {author} {\bibinfo {author} {\bibfnamefont {J.}~\bibnamefont
  {Zhang}}, \bibinfo {author} {\bibfnamefont {X.}~\bibnamefont {Cai}}, \bibinfo
  {author} {\bibfnamefont {W.}~\bibnamefont {Xia}}, \bibinfo {author}
  {\bibfnamefont {A.}~\bibnamefont {Liang}}, \bibinfo {author} {\bibfnamefont
  {J.}~\bibnamefont {Huang}}, \bibinfo {author} {\bibfnamefont
  {C.}~\bibnamefont {Wang}}, \bibinfo {author} {\bibfnamefont {L.}~\bibnamefont
  {Yang}}, \bibinfo {author} {\bibfnamefont {H.}~\bibnamefont {Yuan}}, \bibinfo
  {author} {\bibfnamefont {Y.}~\bibnamefont {Chen}}, \bibinfo {author}
  {\bibfnamefont {S.}~\bibnamefont {Zhang}},  \emph {et~al.},\ }\href
  {https://doi.org/10.1103/PhysRevLett.123.047203} {\bibfield  {journal}
  {\bibinfo  {journal} {Phys. Rev. Lett.}\ }\textbf {\bibinfo {volume} {123}},\
  \bibinfo {pages} {047203} (\bibinfo {year} {2019}{\natexlab{a}})}\BibitemShut
  {NoStop}%
\bibitem [{\citenamefont {Yan}\ \emph {et~al.}(2004)\citenamefont {Yan},
  \citenamefont {Zhou},\ and\ \citenamefont
  {Goodenough}}]{yan2004ferromagnetism}%
  \BibitemOpen
  \bibfield  {author} {\bibinfo {author} {\bibfnamefont {J.~Q.}\ \bibnamefont
  {Yan}}, \bibinfo {author} {\bibfnamefont {J.~S.}\ \bibnamefont {Zhou}}, \
  and\ \bibinfo {author} {\bibfnamefont {J.~B.}\ \bibnamefont {Goodenough}},\
  }\href {https://doi.org/10.1103/PhysRevB.70.014402} {\bibfield  {journal}
  {\bibinfo  {journal} {Phys. Rev. B}\ }\textbf {\bibinfo {volume} {70}},\
  \bibinfo {pages} {014402} (\bibinfo {year} {2004})}\BibitemShut {NoStop}%
\bibitem [{\citenamefont {Mehta}\ \emph {et~al.}(2015)\citenamefont {Mehta},
  \citenamefont {Biskup}, \citenamefont {Jenkins}, \citenamefont {Arenholz},
  \citenamefont {Varela},\ and\ \citenamefont {Suzuki}}]{mehta2015long}%
  \BibitemOpen
  \bibfield  {author} {\bibinfo {author} {\bibfnamefont {V.~V.}\ \bibnamefont
  {Mehta}}, \bibinfo {author} {\bibfnamefont {N.}~\bibnamefont {Biskup}},
  \bibinfo {author} {\bibfnamefont {C.}~\bibnamefont {Jenkins}}, \bibinfo
  {author} {\bibfnamefont {E.}~\bibnamefont {Arenholz}}, \bibinfo {author}
  {\bibfnamefont {M.}~\bibnamefont {Varela}}, \ and\ \bibinfo {author}
  {\bibfnamefont {Y.}~\bibnamefont {Suzuki}},\ }\href
  {https://doi.org/10.1103/PhysRevB.91.144418} {\bibfield  {journal} {\bibinfo
  {journal} {Phys. Rev. B}\ }\textbf {\bibinfo {volume} {91}},\ \bibinfo
  {pages} {144418} (\bibinfo {year} {2015})}\BibitemShut {NoStop}%
\bibitem [{\citenamefont {Zhang}\ \emph
  {et~al.}(2021{\natexlab{a}})\citenamefont {Zhang}, \citenamefont {Gao},
  \citenamefont {Meng}, \citenamefont {Jin}, \citenamefont {Lin}, \citenamefont
  {Wang}, \citenamefont {Xiao}, \citenamefont {Wang}, \citenamefont {Jin},
  \citenamefont {Su} \emph {et~al.}}]{zhang2021near}%
  \BibitemOpen
  \bibfield  {author} {\bibinfo {author} {\bibfnamefont {Q.}~\bibnamefont
  {Zhang}}, \bibinfo {author} {\bibfnamefont {A.}~\bibnamefont {Gao}}, \bibinfo
  {author} {\bibfnamefont {F.}~\bibnamefont {Meng}}, \bibinfo {author}
  {\bibfnamefont {Q.}~\bibnamefont {Jin}}, \bibinfo {author} {\bibfnamefont
  {S.}~\bibnamefont {Lin}}, \bibinfo {author} {\bibfnamefont {X.}~\bibnamefont
  {Wang}}, \bibinfo {author} {\bibfnamefont {D.}~\bibnamefont {Xiao}}, \bibinfo
  {author} {\bibfnamefont {C.}~\bibnamefont {Wang}}, \bibinfo {author}
  {\bibfnamefont {K.-j.}\ \bibnamefont {Jin}}, \bibinfo {author} {\bibfnamefont
  {D.}~\bibnamefont {Su}},  \emph {et~al.},\ }\href
  {https://doi.org/10.1038/s41467-021-22099-y} {\bibfield  {journal} {\bibinfo
  {journal} {Nat. Commun.}\ }\textbf {\bibinfo {volume} {12}},\ \bibinfo
  {pages} {1} (\bibinfo {year} {2021}{\natexlab{a}})}\BibitemShut {NoStop}%
\bibitem [{\citenamefont {Taskin}\ \emph {et~al.}(2003)\citenamefont {Taskin},
  \citenamefont {Lavrov},\ and\ \citenamefont {Ando}}]{taskin2003ising}%
  \BibitemOpen
  \bibfield  {author} {\bibinfo {author} {\bibfnamefont {A.~A.}\ \bibnamefont
  {Taskin}}, \bibinfo {author} {\bibfnamefont {A.~N.}\ \bibnamefont {Lavrov}},
  \ and\ \bibinfo {author} {\bibfnamefont {Y.}~\bibnamefont {Ando}},\ }\href
  {https://doi.org/10.1103/PhysRevLett.90.227201} {\bibfield  {journal}
  {\bibinfo  {journal} {Phys. Rev. Lett.}\ }\textbf {\bibinfo {volume} {90}},\
  \bibinfo {pages} {227201} (\bibinfo {year} {2003})}\BibitemShut {NoStop}%
\bibitem [{\citenamefont {Das}\ \emph {et~al.}(2008)\citenamefont {Das},
  \citenamefont {Waghmare}, \citenamefont {Saha-Dasgupta},\ and\ \citenamefont
  {Sarma}}]{das2008electronic}%
  \BibitemOpen
  \bibfield  {author} {\bibinfo {author} {\bibfnamefont {H.}~\bibnamefont
  {Das}}, \bibinfo {author} {\bibfnamefont {U.~V.}\ \bibnamefont {Waghmare}},
  \bibinfo {author} {\bibfnamefont {T.}~\bibnamefont {Saha-Dasgupta}}, \ and\
  \bibinfo {author} {\bibfnamefont {D.~D.}\ \bibnamefont {Sarma}},\ }\href
  {https://doi.org/10.1103/PhysRevLett.100.186402} {\bibfield  {journal}
  {\bibinfo  {journal} {Phys. Rev. Lett.}\ }\textbf {\bibinfo {volume} {100}},\
  \bibinfo {pages} {186402} (\bibinfo {year} {2008})}\BibitemShut {NoStop}%
\bibitem [{\citenamefont {Feng}\ \emph {et~al.}(2020)\citenamefont {Feng},
  \citenamefont {Deng}, \citenamefont {Segre}, \citenamefont {Croft},
  \citenamefont {Lapidus}, \citenamefont {Frank}, \citenamefont {Shi},
  \citenamefont {Jin}, \citenamefont {Walker},\ and\ \citenamefont
  {Greenblatt}}]{feng2020high}%
  \BibitemOpen
  \bibfield  {author} {\bibinfo {author} {\bibfnamefont {H.~L.}\ \bibnamefont
  {Feng}}, \bibinfo {author} {\bibfnamefont {Z.}~\bibnamefont {Deng}}, \bibinfo
  {author} {\bibfnamefont {C.~U.}\ \bibnamefont {Segre}}, \bibinfo {author}
  {\bibfnamefont {M.}~\bibnamefont {Croft}}, \bibinfo {author} {\bibfnamefont
  {S.~H.}\ \bibnamefont {Lapidus}}, \bibinfo {author} {\bibfnamefont {C.~E.}\
  \bibnamefont {Frank}}, \bibinfo {author} {\bibfnamefont {Y.}~\bibnamefont
  {Shi}}, \bibinfo {author} {\bibfnamefont {C.}~\bibnamefont {Jin}}, \bibinfo
  {author} {\bibfnamefont {D.}~\bibnamefont {Walker}}, \ and\ \bibinfo {author}
  {\bibfnamefont {M.}~\bibnamefont {Greenblatt}},\ }\href
  {https://doi.org/10.1021/acs.inorgchem.0c03402} {\bibfield  {journal}
  {\bibinfo  {journal} {Inorg. Chem.}\ } (\bibinfo {year} {2020})}\BibitemShut
  {NoStop}%
\bibitem [{\citenamefont {McGuire}\ \emph {et~al.}(2015)\citenamefont
  {McGuire}, \citenamefont {Dixit}, \citenamefont {Cooper},\ and\ \citenamefont
  {Sales}}]{mcguire2015coupling}%
  \BibitemOpen
  \bibfield  {author} {\bibinfo {author} {\bibfnamefont {M.~A.}\ \bibnamefont
  {McGuire}}, \bibinfo {author} {\bibfnamefont {H.}~\bibnamefont {Dixit}},
  \bibinfo {author} {\bibfnamefont {V.~R.}\ \bibnamefont {Cooper}}, \ and\
  \bibinfo {author} {\bibfnamefont {B.~C.}\ \bibnamefont {Sales}},\ }\href
  {https://doi.org/10.1021/cm504242t} {\bibfield  {journal} {\bibinfo
  {journal} {Chem. Mater.}\ }\textbf {\bibinfo {volume} {27}},\ \bibinfo
  {pages} {612} (\bibinfo {year} {2015})}\BibitemShut {NoStop}%
\bibitem [{\citenamefont {Gong}\ \emph {et~al.}(2017)\citenamefont {Gong},
  \citenamefont {Li}, \citenamefont {Li}, \citenamefont {Ji}, \citenamefont
  {Stern}, \citenamefont {Xia}, \citenamefont {Cao}, \citenamefont {Bao},
  \citenamefont {Wang}, \citenamefont {Wang} \emph
  {et~al.}}]{gong2017discovery}%
  \BibitemOpen
  \bibfield  {author} {\bibinfo {author} {\bibfnamefont {C.}~\bibnamefont
  {Gong}}, \bibinfo {author} {\bibfnamefont {L.}~\bibnamefont {Li}}, \bibinfo
  {author} {\bibfnamefont {Z.}~\bibnamefont {Li}}, \bibinfo {author}
  {\bibfnamefont {H.}~\bibnamefont {Ji}}, \bibinfo {author} {\bibfnamefont
  {A.}~\bibnamefont {Stern}}, \bibinfo {author} {\bibfnamefont
  {Y.}~\bibnamefont {Xia}}, \bibinfo {author} {\bibfnamefont {T.}~\bibnamefont
  {Cao}}, \bibinfo {author} {\bibfnamefont {W.}~\bibnamefont {Bao}}, \bibinfo
  {author} {\bibfnamefont {C.}~\bibnamefont {Wang}}, \bibinfo {author}
  {\bibfnamefont {Y.}~\bibnamefont {Wang}},  \emph {et~al.},\ }\href
  {https://doi.org/10.1038/nature22060} {\bibfield  {journal} {\bibinfo
  {journal} {Nature (London)}\ }\textbf {\bibinfo {volume} {546}},\ \bibinfo
  {pages} {265} (\bibinfo {year} {2017})}\BibitemShut {NoStop}%
\bibitem [{\citenamefont {Huang}\ \emph {et~al.}(2020)\citenamefont {Huang},
  \citenamefont {Liu}, \citenamefont {Mansikkam{\"a}ki}, \citenamefont {Vieru},
  \citenamefont {Iwahara},\ and\ \citenamefont
  {Chibotaru}}]{huang2020ferromagnetic}%
  \BibitemOpen
  \bibfield  {author} {\bibinfo {author} {\bibfnamefont {Z.}~\bibnamefont
  {Huang}}, \bibinfo {author} {\bibfnamefont {D.}~\bibnamefont {Liu}}, \bibinfo
  {author} {\bibfnamefont {A.}~\bibnamefont {Mansikkam{\"a}ki}}, \bibinfo
  {author} {\bibfnamefont {V.}~\bibnamefont {Vieru}}, \bibinfo {author}
  {\bibfnamefont {N.}~\bibnamefont {Iwahara}}, \ and\ \bibinfo {author}
  {\bibfnamefont {L.~F.}\ \bibnamefont {Chibotaru}},\ }\href
  {https://doi.org/10.1103/PhysRevResearch.2.033430} {\bibfield  {journal}
  {\bibinfo  {journal} {Phys. Rev. Research}\ }\textbf {\bibinfo {volume}
  {2}},\ \bibinfo {pages} {033430} (\bibinfo {year} {2020})}\BibitemShut
  {NoStop}%
\bibitem [{\citenamefont {Steiner}\ \emph {et~al.}(1976)\citenamefont
  {Steiner}, \citenamefont {Villain},\ and\ \citenamefont
  {Windsor}}]{steiner1976theoretical}%
  \BibitemOpen
  \bibfield  {author} {\bibinfo {author} {\bibfnamefont {M.}~\bibnamefont
  {Steiner}}, \bibinfo {author} {\bibfnamefont {J.}~\bibnamefont {Villain}}, \
  and\ \bibinfo {author} {\bibfnamefont {C.}~\bibnamefont {Windsor}},\ }\href
  {https://doi.org/10.1080/00018737600101372} {\bibfield  {journal} {\bibinfo
  {journal} {Adv. Phys.}\ }\textbf {\bibinfo {volume} {25}},\ \bibinfo {pages}
  {87} (\bibinfo {year} {1976})}\BibitemShut {NoStop}%
\bibitem [{\citenamefont {Anderson}(1950)}]{anderson1950antiferromagnetism}%
  \BibitemOpen
  \bibfield  {author} {\bibinfo {author} {\bibfnamefont {P.~W.}\ \bibnamefont
  {Anderson}},\ }\href {https://doi.org/10.1103/PhysRev.79.350} {\bibfield
  {journal} {\bibinfo  {journal} {Phys. Rev.}\ }\textbf {\bibinfo {volume}
  {79}},\ \bibinfo {pages} {350} (\bibinfo {year} {1950})}\BibitemShut
  {NoStop}%
\bibitem [{\citenamefont {Goodenough}(1955)}]{goodenough1955theory}%
  \BibitemOpen
  \bibfield  {author} {\bibinfo {author} {\bibfnamefont {J.~B.}\ \bibnamefont
  {Goodenough}},\ }\href {https://doi.org/10.1103/PhysRev.100.564} {\bibfield
  {journal} {\bibinfo  {journal} {Phys. Rev.}\ }\textbf {\bibinfo {volume}
  {100}},\ \bibinfo {pages} {564} (\bibinfo {year} {1955})}\BibitemShut
  {NoStop}%
\bibitem [{\citenamefont {Goodenough}(1958)}]{goodenough1958interpretation}%
  \BibitemOpen
  \bibfield  {author} {\bibinfo {author} {\bibfnamefont {J.~B.}\ \bibnamefont
  {Goodenough}},\ }\href {https://doi.org/10.1016/0022-3697(58)90107-0}
  {\bibfield  {journal} {\bibinfo  {journal} {J. Phys. Chem. Solids}\ }\textbf
  {\bibinfo {volume} {6}},\ \bibinfo {pages} {287} (\bibinfo {year}
  {1958})}\BibitemShut {NoStop}%
\bibitem [{\citenamefont {Kanamori}(1959)}]{kanamori1959superexchange}%
  \BibitemOpen
  \bibfield  {author} {\bibinfo {author} {\bibfnamefont {J.}~\bibnamefont
  {Kanamori}},\ }\href {https://doi.org/10.1016/0022-3697(59)90061-7}
  {\bibfield  {journal} {\bibinfo  {journal} {J. Phys. Chem. Solids}\ }\textbf
  {\bibinfo {volume} {10}},\ \bibinfo {pages} {87} (\bibinfo {year}
  {1959})}\BibitemShut {NoStop}%
\bibitem [{thr()}]{threemechanism}%
  \BibitemOpen
  \href@noop {} {}\bibinfo {note} {The present study suggests that COFS host
  the competition between mechanism (a) and (c). Interestingly, the abnormally
  large {\it inter-orbital hopping} amplitude together with large $J_{\rm H}$
  allows the FM kinetic mechanism (c) to dominate over (a).}\BibitemShut
  {Stop}%
\bibitem [{\citenamefont {Kuroki}\ \emph {et~al.}(2008)\citenamefont {Kuroki},
  \citenamefont {Onari}, \citenamefont {Arita}, \citenamefont {Usui},
  \citenamefont {Tanaka}, \citenamefont {Kontani},\ and\ \citenamefont
  {Aoki}}]{kuroki2008unconventional}%
  \BibitemOpen
  \bibfield  {author} {\bibinfo {author} {\bibfnamefont {K.}~\bibnamefont
  {Kuroki}}, \bibinfo {author} {\bibfnamefont {S.}~\bibnamefont {Onari}},
  \bibinfo {author} {\bibfnamefont {R.}~\bibnamefont {Arita}}, \bibinfo
  {author} {\bibfnamefont {H.}~\bibnamefont {Usui}}, \bibinfo {author}
  {\bibfnamefont {Y.}~\bibnamefont {Tanaka}}, \bibinfo {author} {\bibfnamefont
  {H.}~\bibnamefont {Kontani}}, \ and\ \bibinfo {author} {\bibfnamefont
  {H.}~\bibnamefont {Aoki}},\ }\href
  {https://doi.org/10.1103/PhysRevLett.101.087004} {\bibfield  {journal}
  {\bibinfo  {journal} {Phys. Rev. Lett.}\ }\textbf {\bibinfo {volume} {101}},\
  \bibinfo {pages} {087004} (\bibinfo {year} {2008})}\BibitemShut {NoStop}%
\bibitem [{\citenamefont {Pandey}\ \emph {et~al.}(2021)\citenamefont {Pandey},
  \citenamefont {Zhang}, \citenamefont {Kaushal}, \citenamefont {Soni},
  \citenamefont {Lin}, \citenamefont {Hu}, \citenamefont {Alvarez},\ and\
  \citenamefont {Dagotto}}]{pandey2021origin}%
  \BibitemOpen
  \bibfield  {author} {\bibinfo {author} {\bibfnamefont {B.}~\bibnamefont
  {Pandey}}, \bibinfo {author} {\bibfnamefont {Y.}~\bibnamefont {Zhang}},
  \bibinfo {author} {\bibfnamefont {N.}~\bibnamefont {Kaushal}}, \bibinfo
  {author} {\bibfnamefont {R.}~\bibnamefont {Soni}}, \bibinfo {author}
  {\bibfnamefont {L.-F.}\ \bibnamefont {Lin}}, \bibinfo {author} {\bibfnamefont
  {W.-J.}\ \bibnamefont {Hu}}, \bibinfo {author} {\bibfnamefont
  {G.}~\bibnamefont {Alvarez}}, \ and\ \bibinfo {author} {\bibfnamefont
  {E.}~\bibnamefont {Dagotto}},\ }\href
  {https://doi.org/10.1103/PhysRevB.103.045115} {\bibfield  {journal} {\bibinfo
   {journal} {Phys. Rev. B}\ }\textbf {\bibinfo {volume} {103}},\ \bibinfo
  {pages} {045115} (\bibinfo {year} {2021})}\BibitemShut {NoStop}%
\bibitem [{\citenamefont {Lin}\ \emph {et~al.}(2021{\natexlab{b}})\citenamefont
  {Lin}, \citenamefont {Kaushal}, \citenamefont {Zhang}, \citenamefont
  {Moreo},\ and\ \citenamefont {Dagotto}}]{lin2021orbital}%
  \BibitemOpen
  \bibfield  {author} {\bibinfo {author} {\bibfnamefont {L.-F.}\ \bibnamefont
  {Lin}}, \bibinfo {author} {\bibfnamefont {N.}~\bibnamefont {Kaushal}},
  \bibinfo {author} {\bibfnamefont {Y.}~\bibnamefont {Zhang}}, \bibinfo
  {author} {\bibfnamefont {A.}~\bibnamefont {Moreo}}, \ and\ \bibinfo {author}
  {\bibfnamefont {E.}~\bibnamefont {Dagotto}},\ }\href
  {https://doi.org/10.1103/PhysRevMaterials.5.025001} {\bibfield  {journal}
  {\bibinfo  {journal} {Phys. Rev. Mater.}\ }\textbf {\bibinfo {volume} {5}},\
  \bibinfo {pages} {025001} (\bibinfo {year} {2021}{\natexlab{b}})}\BibitemShut
  {NoStop}%
\bibitem [{\citenamefont {Lin}\ \emph {et~al.}(2019)\citenamefont {Lin},
  \citenamefont {Zhang}, \citenamefont {Moreo}, \citenamefont {Dagotto},\ and\
  \citenamefont {Dong}}]{lin2019frustrated}%
  \BibitemOpen
  \bibfield  {author} {\bibinfo {author} {\bibfnamefont {L.-F.}\ \bibnamefont
  {Lin}}, \bibinfo {author} {\bibfnamefont {Y.}~\bibnamefont {Zhang}}, \bibinfo
  {author} {\bibfnamefont {A.}~\bibnamefont {Moreo}}, \bibinfo {author}
  {\bibfnamefont {E.}~\bibnamefont {Dagotto}}, \ and\ \bibinfo {author}
  {\bibfnamefont {S.}~\bibnamefont {Dong}},\ }\href
  {https://doi.org/10.1103/PhysRevLett.123.067601} {\bibfield  {journal}
  {\bibinfo  {journal} {Phys. Rev. Lett.}\ }\textbf {\bibinfo {volume} {123}},\
  \bibinfo {pages} {067601} (\bibinfo {year} {2019})}\BibitemShut {NoStop}%
\bibitem [{\citenamefont {Zhang}\ \emph
  {et~al.}(2021{\natexlab{b}})\citenamefont {Zhang}, \citenamefont {Lin},
  \citenamefont {Moreo}, \citenamefont {Alvarez},\ and\ \citenamefont
  {Dagotto}}]{zhang2021peierls}%
  \BibitemOpen
  \bibfield  {author} {\bibinfo {author} {\bibfnamefont {Y.}~\bibnamefont
  {Zhang}}, \bibinfo {author} {\bibfnamefont {L.-F.}\ \bibnamefont {Lin}},
  \bibinfo {author} {\bibfnamefont {A.}~\bibnamefont {Moreo}}, \bibinfo
  {author} {\bibfnamefont {G.}~\bibnamefont {Alvarez}}, \ and\ \bibinfo
  {author} {\bibfnamefont {E.}~\bibnamefont {Dagotto}},\ }\href
  {https://doi.org/10.1103/PhysRevB.103.L121114} {\bibfield  {journal}
  {\bibinfo  {journal} {Phys. Rev. B}\ }\textbf {\bibinfo {volume} {103}},\
  \bibinfo {pages} {L121114} (\bibinfo {year}
  {2021}{\natexlab{b}})}\BibitemShut {NoStop}%
\bibitem [{\citenamefont {Zhang}\ \emph
  {et~al.}(2019{\natexlab{b}})\citenamefont {Zhang}, \citenamefont {Lin},
  \citenamefont {Moreo}, \citenamefont {Dong},\ and\ \citenamefont
  {Dagotto}}]{zhang2019magnetic}%
  \BibitemOpen
  \bibfield  {author} {\bibinfo {author} {\bibfnamefont {Y.}~\bibnamefont
  {Zhang}}, \bibinfo {author} {\bibfnamefont {L.-F.}\ \bibnamefont {Lin}},
  \bibinfo {author} {\bibfnamefont {A.}~\bibnamefont {Moreo}}, \bibinfo
  {author} {\bibfnamefont {S.}~\bibnamefont {Dong}}, \ and\ \bibinfo {author}
  {\bibfnamefont {E.}~\bibnamefont {Dagotto}},\ }\href
  {https://doi.org/10.1103/PhysRevB.100.184419} {\bibfield  {journal} {\bibinfo
   {journal} {Phys. Rev. B}\ }\textbf {\bibinfo {volume} {100}},\ \bibinfo
  {pages} {184419} (\bibinfo {year} {2019}{\natexlab{b}})}\BibitemShut
  {NoStop}%
\bibitem [{\citenamefont {Zhang}\ \emph
  {et~al.}(2020{\natexlab{a}})\citenamefont {Zhang}, \citenamefont {Lin},
  \citenamefont {Moreo}, \citenamefont {Dong},\ and\ \citenamefont
  {Dagotto}}]{zhang2020iron}%
  \BibitemOpen
  \bibfield  {author} {\bibinfo {author} {\bibfnamefont {Y.}~\bibnamefont
  {Zhang}}, \bibinfo {author} {\bibfnamefont {L.-F.}\ \bibnamefont {Lin}},
  \bibinfo {author} {\bibfnamefont {A.}~\bibnamefont {Moreo}}, \bibinfo
  {author} {\bibfnamefont {S.}~\bibnamefont {Dong}}, \ and\ \bibinfo {author}
  {\bibfnamefont {E.}~\bibnamefont {Dagotto}},\ }\href
  {https://doi.org/10.1103/PhysRevB.101.144417} {\bibfield  {journal} {\bibinfo
   {journal} {Phys. Rev. B}\ }\textbf {\bibinfo {volume} {101}},\ \bibinfo
  {pages} {144417} (\bibinfo {year} {2020}{\natexlab{a}})}\BibitemShut
  {NoStop}%
\bibitem [{\citenamefont {Zhang}\ \emph
  {et~al.}(2020{\natexlab{b}})\citenamefont {Zhang}, \citenamefont {Lin},
  \citenamefont {Moreo}, \citenamefont {Dong},\ and\ \citenamefont
  {Dagotto}}]{zhang2020first}%
  \BibitemOpen
  \bibfield  {author} {\bibinfo {author} {\bibfnamefont {Y.}~\bibnamefont
  {Zhang}}, \bibinfo {author} {\bibfnamefont {L.-F.}\ \bibnamefont {Lin}},
  \bibinfo {author} {\bibfnamefont {A.}~\bibnamefont {Moreo}}, \bibinfo
  {author} {\bibfnamefont {S.}~\bibnamefont {Dong}}, \ and\ \bibinfo {author}
  {\bibfnamefont {E.}~\bibnamefont {Dagotto}},\ }\href
  {https://doi.org/10.1103/PhysRevB.101.174106} {\bibfield  {journal} {\bibinfo
   {journal} {Phys. Rev. B}\ }\textbf {\bibinfo {volume} {101}},\ \bibinfo
  {pages} {174106} (\bibinfo {year} {2020}{\natexlab{b}})}\BibitemShut
  {NoStop}%
\bibitem [{\citenamefont {Herbrych}\ \emph {et~al.}(2020)\citenamefont
  {Herbrych}, \citenamefont {Heverhagen}, \citenamefont {Alvarez},
  \citenamefont {Daghofer}, \citenamefont {Moreo},\ and\ \citenamefont
  {Dagotto}}]{herbrych2020block}%
  \BibitemOpen
  \bibfield  {author} {\bibinfo {author} {\bibfnamefont {J.}~\bibnamefont
  {Herbrych}}, \bibinfo {author} {\bibfnamefont {J.}~\bibnamefont
  {Heverhagen}}, \bibinfo {author} {\bibfnamefont {G.}~\bibnamefont {Alvarez}},
  \bibinfo {author} {\bibfnamefont {M.}~\bibnamefont {Daghofer}}, \bibinfo
  {author} {\bibfnamefont {A.}~\bibnamefont {Moreo}}, \ and\ \bibinfo {author}
  {\bibfnamefont {E.}~\bibnamefont {Dagotto}},\ }\href
  {https://doi.org/10.1073/pnas.2001141117} {\bibfield  {journal} {\bibinfo
  {journal} {Proc. Natl. Acad. Sci. USA}\ }\textbf {\bibinfo {volume} {117}},\
  \bibinfo {pages} {16226} (\bibinfo {year} {2020})}\BibitemShut {NoStop}%
\bibitem [{\citenamefont {Pandey}\ \emph {et~al.}(2020)\citenamefont {Pandey},
  \citenamefont {Lin}, \citenamefont {Soni}, \citenamefont {Kaushal},
  \citenamefont {Herbrych}, \citenamefont {Alvarez},\ and\ \citenamefont
  {Dagotto}}]{pandey2020prediction}%
  \BibitemOpen
  \bibfield  {author} {\bibinfo {author} {\bibfnamefont {B.}~\bibnamefont
  {Pandey}}, \bibinfo {author} {\bibfnamefont {L.-F.}\ \bibnamefont {Lin}},
  \bibinfo {author} {\bibfnamefont {R.}~\bibnamefont {Soni}}, \bibinfo {author}
  {\bibfnamefont {N.}~\bibnamefont {Kaushal}}, \bibinfo {author} {\bibfnamefont
  {J.}~\bibnamefont {Herbrych}}, \bibinfo {author} {\bibfnamefont
  {G.}~\bibnamefont {Alvarez}}, \ and\ \bibinfo {author} {\bibfnamefont
  {E.}~\bibnamefont {Dagotto}},\ }\href
  {https://doi.org/10.1103/PhysRevB.102.035149} {\bibfield  {journal} {\bibinfo
   {journal} {Phys. Rev. B}\ }\textbf {\bibinfo {volume} {102}},\ \bibinfo
  {pages} {035149} (\bibinfo {year} {2020})}\BibitemShut {NoStop}%
\bibitem [{\citenamefont {Patel}\ \emph {et~al.}(2019)\citenamefont {Patel},
  \citenamefont {Nocera}, \citenamefont {Alvarez}, \citenamefont {Moreo},
  \citenamefont {Johnston},\ and\ \citenamefont
  {Dagotto}}]{patel2019fingerprints}%
  \BibitemOpen
  \bibfield  {author} {\bibinfo {author} {\bibfnamefont {N.}~\bibnamefont
  {Patel}}, \bibinfo {author} {\bibfnamefont {A.}~\bibnamefont {Nocera}},
  \bibinfo {author} {\bibfnamefont {G.}~\bibnamefont {Alvarez}}, \bibinfo
  {author} {\bibfnamefont {A.}~\bibnamefont {Moreo}}, \bibinfo {author}
  {\bibfnamefont {S.}~\bibnamefont {Johnston}}, \ and\ \bibinfo {author}
  {\bibfnamefont {E.}~\bibnamefont {Dagotto}},\ }\href
  {https://doi.org/10.1038/s42005-019-0155-3} {\bibfield  {journal} {\bibinfo
  {journal} {Comm. Phys.}\ }\textbf {\bibinfo {volume} {2}},\ \bibinfo {pages}
  {64} (\bibinfo {year} {2019})}\BibitemShut {NoStop}%
\bibitem [{\citenamefont {Herbrych}\ \emph {et~al.}(2019)\citenamefont
  {Herbrych}, \citenamefont {Heverhagen}, \citenamefont {Patel}, \citenamefont
  {Alvarez}, \citenamefont {Daghofer}, \citenamefont {Moreo},\ and\
  \citenamefont {Dagotto}}]{herbrych2019novel}%
  \BibitemOpen
  \bibfield  {author} {\bibinfo {author} {\bibfnamefont {J.}~\bibnamefont
  {Herbrych}}, \bibinfo {author} {\bibfnamefont {J.}~\bibnamefont
  {Heverhagen}}, \bibinfo {author} {\bibfnamefont {N.~D.}\ \bibnamefont
  {Patel}}, \bibinfo {author} {\bibfnamefont {G.}~\bibnamefont {Alvarez}},
  \bibinfo {author} {\bibfnamefont {M.}~\bibnamefont {Daghofer}}, \bibinfo
  {author} {\bibfnamefont {A.}~\bibnamefont {Moreo}}, \ and\ \bibinfo {author}
  {\bibfnamefont {E.}~\bibnamefont {Dagotto}},\ }\href
  {https://doi.org/10.1103/PhysRevLett.123.027203} {\bibfield  {journal}
  {\bibinfo  {journal} {Phys. Rev. Lett.}\ }\textbf {\bibinfo {volume} {123}},\
  \bibinfo {pages} {027203} (\bibinfo {year} {2019})}\BibitemShut {NoStop}%
\bibitem [{\citenamefont {McCabe}\ \emph {et~al.}(2011)\citenamefont {McCabe},
  \citenamefont {Free},\ and\ \citenamefont {Evans}}]{mccabe2011new}%
  \BibitemOpen
  \bibfield  {author} {\bibinfo {author} {\bibfnamefont {E.~E.}\ \bibnamefont
  {McCabe}}, \bibinfo {author} {\bibfnamefont {D.~G.}\ \bibnamefont {Free}}, \
  and\ \bibinfo {author} {\bibfnamefont {J.~S.}\ \bibnamefont {Evans}},\ }\href
  {https://doi.org/10.1039/c0cc03477k} {\bibfield  {journal} {\bibinfo
  {journal} {Chem. Comm.}\ }\textbf {\bibinfo {volume} {47}},\ \bibinfo {pages}
  {1261} (\bibinfo {year} {2011})}\BibitemShut {NoStop}%
\bibitem [{\citenamefont {McCabe}\ \emph {et~al.}(2014)\citenamefont {McCabe},
  \citenamefont {Stock}, \citenamefont {Bettis}, \citenamefont {Whangbo},\ and\
  \citenamefont {Evans}}]{mccabe2014magnetism}%
  \BibitemOpen
  \bibfield  {author} {\bibinfo {author} {\bibfnamefont {E.~E.}\ \bibnamefont
  {McCabe}}, \bibinfo {author} {\bibfnamefont {C.}~\bibnamefont {Stock}},
  \bibinfo {author} {\bibfnamefont {J.~L.}\ \bibnamefont {Bettis}}, \bibinfo
  {author} {\bibfnamefont {M.~H.}\ \bibnamefont {Whangbo}}, \ and\ \bibinfo
  {author} {\bibfnamefont {J.~S.~O.}\ \bibnamefont {Evans}},\ }\href
  {https://doi.org/10.1103/PhysRevB.90.235115} {\bibfield  {journal} {\bibinfo
  {journal} {Phys. Rev. B}\ }\textbf {\bibinfo {volume} {90}},\ \bibinfo
  {pages} {235115} (\bibinfo {year} {2014})}\BibitemShut {NoStop}%
\bibitem [{\citenamefont {Kresse}\ and\ \citenamefont
  {Joubert}(1999)}]{Kresse:Prb99}%
  \BibitemOpen
  \bibfield  {author} {\bibinfo {author} {\bibfnamefont {G.}~\bibnamefont
  {Kresse}}\ and\ \bibinfo {author} {\bibfnamefont {D.}~\bibnamefont
  {Joubert}},\ }\href {https://doi.org/10.1103/PhysRevB.59.1758} {\bibfield
  {journal} {\bibinfo  {journal} {Phys. Rev. B}\ }\textbf {\bibinfo {volume}
  {59}},\ \bibinfo {pages} {1758} (\bibinfo {year} {1999})}\BibitemShut
  {NoStop}%
\bibitem [{\citenamefont {Bl\"{o}chl}(1994)}]{Blochl:Prb2}%
  \BibitemOpen
  \bibfield  {author} {\bibinfo {author} {\bibfnamefont {P.~E.}\ \bibnamefont
  {Bl\"{o}chl}},\ }\href {https://doi.org/10.1103/PhysRevB.50.17953} {\bibfield
   {journal} {\bibinfo  {journal} {Phys. Rev. B}\ }\textbf {\bibinfo {volume}
  {50}},\ \bibinfo {pages} {17953} (\bibinfo {year} {1994})}\BibitemShut
  {NoStop}%
\bibitem [{\citenamefont {Perdew}\ \emph {et~al.}(2008)\citenamefont {Perdew},
  \citenamefont {Ruzsinszky}, \citenamefont {Csonka}, \citenamefont {Vydrov},
  \citenamefont {Scuseria}, \citenamefont {Constantin}, \citenamefont {Zhou},\
  and\ \citenamefont {Burke}}]{Perdew:Prl08}%
  \BibitemOpen
  \bibfield  {author} {\bibinfo {author} {\bibfnamefont {J.~P.}\ \bibnamefont
  {Perdew}}, \bibinfo {author} {\bibfnamefont {A.}~\bibnamefont {Ruzsinszky}},
  \bibinfo {author} {\bibfnamefont {G.~I.}\ \bibnamefont {Csonka}}, \bibinfo
  {author} {\bibfnamefont {O.~A.}\ \bibnamefont {Vydrov}}, \bibinfo {author}
  {\bibfnamefont {G.~E.}\ \bibnamefont {Scuseria}}, \bibinfo {author}
  {\bibfnamefont {L.~A.}\ \bibnamefont {Constantin}}, \bibinfo {author}
  {\bibfnamefont {X.}~\bibnamefont {Zhou}}, \ and\ \bibinfo {author}
  {\bibfnamefont {K.}~\bibnamefont {Burke}},\ }\href
  {https://doi.org/10.1103/PhysRevLett.100.136406} {\bibfield  {journal}
  {\bibinfo  {journal} {Phys. Rev. Lett.}\ }\textbf {\bibinfo {volume} {100}},\
  \bibinfo {pages} {136406} (\bibinfo {year} {2008})}\BibitemShut {NoStop}%
\bibitem [{\citenamefont {Marzari}\ and\ \citenamefont
  {Vanderbilt}(1997)}]{marzari1997maximally}%
  \BibitemOpen
  \bibfield  {author} {\bibinfo {author} {\bibfnamefont {N.}~\bibnamefont
  {Marzari}}\ and\ \bibinfo {author} {\bibfnamefont {D.}~\bibnamefont
  {Vanderbilt}},\ }\href {https://doi.org/10.1103/PhysRevB.56.12847} {\bibfield
   {journal} {\bibinfo  {journal} {Phys. Rev. B}\ }\textbf {\bibinfo {volume}
  {56}},\ \bibinfo {pages} {12847} (\bibinfo {year} {1997})}\BibitemShut
  {NoStop}%
\bibitem [{\citenamefont {Mostofi}\ \emph {et~al.}(2008)\citenamefont
  {Mostofi}, \citenamefont {Yates}, \citenamefont {Lee}, \citenamefont {Souza},
  \citenamefont {Vanderbilt},\ and\ \citenamefont
  {Marzari}}]{mostofi2008wannier90}%
  \BibitemOpen
  \bibfield  {author} {\bibinfo {author} {\bibfnamefont {A.~A.}\ \bibnamefont
  {Mostofi}}, \bibinfo {author} {\bibfnamefont {J.~R.}\ \bibnamefont {Yates}},
  \bibinfo {author} {\bibfnamefont {Y.-S.}\ \bibnamefont {Lee}}, \bibinfo
  {author} {\bibfnamefont {I.}~\bibnamefont {Souza}}, \bibinfo {author}
  {\bibfnamefont {D.}~\bibnamefont {Vanderbilt}}, \ and\ \bibinfo {author}
  {\bibfnamefont {N.}~\bibnamefont {Marzari}},\ }\href
  {https://doi.org/10.1016/j.cpc.2007.11.016} {\bibfield  {journal} {\bibinfo
  {journal} {Comput. Phys. Commun.}\ }\textbf {\bibinfo {volume} {178}},\
  \bibinfo {pages} {685} (\bibinfo {year} {2008})}\BibitemShut {NoStop}%
\bibitem [{com()}]{comments_hoppings}%
  \BibitemOpen
  \href@noop {} {}\bibinfo {note} {Note that we calculate hoppings and crystal
  fields based on DFT without including interaction effects, as most commonly
  done in the literature. Alternatively, we could have calculated these
  parameters using DFT in the FM spin background relevant for COFS. But this
  procedure would have not allowed us to provide the entire phase diagram,
  including non-FM phases, as shown in the DMRG section. Another interesting
  observation is that in Fig.2 SM, we noticed that even at very large Hubbard
  coupling the bandwidth of, e.g., the lower and upper Hubbard bands remain
  similar in value to the original non-interacting bandwidth. Thus, contrary to
  naive expectation, the bands do not flatten at large $U$ although they seem
  flat when compared to the large gaps generated by $U$.}\BibitemShut {Stop}%
\bibitem [{Foo()}]{Footnote}%
  \BibitemOpen
  \href@noop {} {}\bibinfo {note} {Hubbard-like model commonly discards the FM
  direct exchange as mentioned by Wei {\it et. al.} \cite{ku2002insulating},
  but direct exchange could be legitimately omitted in our case since the
  kinetic exchange dominates over the direct exchange.}\BibitemShut {Stop}%
\bibitem [{Sup()}]{Supp}%
  \BibitemOpen
  \href@noop {} {}\bibinfo {note} {See Supplemental Material for more details
  of the method, hopping matrix, and spectrum \& DOS plot, which includes Refs.
  ~\cite{Kresse:Prb99,Blochl:Prb2,Perdew:Prl08,mccabe2011new,mccabe2014magnetism,marzari1997maximally,mostofi2008wannier90,momma2011vesta,white1992density,white1993density,schollwock2005density,hallberg2006new,alvarez2009density}.}\BibitemShut
  {Stop}%
\bibitem [{\citenamefont {White}(1992)}]{white1992density}%
  \BibitemOpen
  \bibfield  {author} {\bibinfo {author} {\bibfnamefont {S.~R.}\ \bibnamefont
  {White}},\ }\href {https://doi.org/10.1103/PhysRevLett.69.2863} {\bibfield
  {journal} {\bibinfo  {journal} {Phys. Rev. Lett.}\ }\textbf {\bibinfo
  {volume} {69}},\ \bibinfo {pages} {2863} (\bibinfo {year}
  {1992})}\BibitemShut {NoStop}%
\bibitem [{\citenamefont {White}(1993)}]{white1993density}%
  \BibitemOpen
  \bibfield  {author} {\bibinfo {author} {\bibfnamefont {S.~R.}\ \bibnamefont
  {White}},\ }\href {https://doi.org/10.1103/PhysRevB.48.10345} {\bibfield
  {journal} {\bibinfo  {journal} {Phys. Rev. B}\ }\textbf {\bibinfo {volume}
  {48}},\ \bibinfo {pages} {10345} (\bibinfo {year} {1993})}\BibitemShut
  {NoStop}%
\bibitem [{\citenamefont {Schollw{\"o}ck}(2005)}]{schollwock2005density}%
  \BibitemOpen
  \bibfield  {author} {\bibinfo {author} {\bibfnamefont {U.}~\bibnamefont
  {Schollw{\"o}ck}},\ }\href {https://doi.org/10.1103/RevModPhys.77.259}
  {\bibfield  {journal} {\bibinfo  {journal} {Rev. Mod. Phys.}\ }\textbf
  {\bibinfo {volume} {77}},\ \bibinfo {pages} {259} (\bibinfo {year}
  {2005})}\BibitemShut {NoStop}%
\bibitem [{\citenamefont {Hallberg}(2006)}]{hallberg2006new}%
  \BibitemOpen
  \bibfield  {author} {\bibinfo {author} {\bibfnamefont {K.~A.}\ \bibnamefont
  {Hallberg}},\ }\href {https://doi.org/10.1080/00018730600766432} {\bibfield
  {journal} {\bibinfo  {journal} {Adv. Phys.}\ }\textbf {\bibinfo {volume}
  {55}},\ \bibinfo {pages} {477} (\bibinfo {year} {2006})}\BibitemShut
  {NoStop}%
\bibitem [{\citenamefont {Alvarez}(2009)}]{alvarez2009density}%
  \BibitemOpen
  \bibfield  {author} {\bibinfo {author} {\bibfnamefont {G.}~\bibnamefont
  {Alvarez}},\ }\href {https://doi.org/10.1016/j.cpc.2009.02.016} {\bibfield
  {journal} {\bibinfo  {journal} {Comput. Phys. Commun.}\ }\textbf {\bibinfo
  {volume} {180}},\ \bibinfo {pages} {1572} (\bibinfo {year}
  {2009})}\BibitemShut {NoStop}%
\bibitem [{\citenamefont {Yu}\ \emph {et~al.}(2021)\citenamefont {Yu},
  \citenamefont {Hu}, \citenamefont {Nica}, \citenamefont {Zhu},\ and\
  \citenamefont {Si}}]{yu2020orbital}%
  \BibitemOpen
  \bibfield  {author} {\bibinfo {author} {\bibfnamefont {R.}~\bibnamefont
  {Yu}}, \bibinfo {author} {\bibfnamefont {H.}~\bibnamefont {Hu}}, \bibinfo
  {author} {\bibfnamefont {E.~M.}\ \bibnamefont {Nica}}, \bibinfo {author}
  {\bibfnamefont {J.-X.}\ \bibnamefont {Zhu}}, \ and\ \bibinfo {author}
  {\bibfnamefont {Q.}~\bibnamefont {Si}},\ }\href
  {https://doi.org/10.3389/fphy.2021.578347} {\bibfield  {journal} {\bibinfo
  {journal} {Front. Phys.}\ }\textbf {\bibinfo {volume} {9}},\ \bibinfo {pages}
  {578347} (\bibinfo {year} {2021})}\BibitemShut {NoStop}%
\bibitem [{\citenamefont {Rinc{\'o}n}\ \emph {et~al.}(2014)\citenamefont
  {Rinc{\'o}n}, \citenamefont {Moreo}, \citenamefont {Alvarez},\ and\
  \citenamefont {Dagotto}}]{rincon2014exotic}%
  \BibitemOpen
  \bibfield  {author} {\bibinfo {author} {\bibfnamefont {J.}~\bibnamefont
  {Rinc{\'o}n}}, \bibinfo {author} {\bibfnamefont {A.}~\bibnamefont {Moreo}},
  \bibinfo {author} {\bibfnamefont {G.}~\bibnamefont {Alvarez}}, \ and\
  \bibinfo {author} {\bibfnamefont {E.}~\bibnamefont {Dagotto}},\ }\href
  {https://doi.org/10.1103/PhysRevLett.112.106405} {\bibfield  {journal}
  {\bibinfo  {journal} {Phys. Rev. Lett.}\ }\textbf {\bibinfo {volume} {112}},\
  \bibinfo {pages} {106405} (\bibinfo {year} {2014})}\BibitemShut {NoStop}%
\bibitem [{mom()}]{moment}%
  \BibitemOpen
  \href@noop {} {}\bibinfo {note} {The total spin square equals 2, which means
  \textbf{S}=1 and spin moment is $2~\mu_{\rm B}$ for 3 orbitals, in agreement
  with the high spin state observed by experiment. If we consider the 5
  orbitals, then the final spin magnetic moment can be roughly estimated as
  $4~\mu_{\rm B}$, which is slightly larger but close to the experimental
  observed value $\sim 3.14-3.33~\mu_{\rm
  B}$~\cite{mccabe2011new,mccabe2014magnetism}.}\BibitemShut {Stop}%
\bibitem [{\citenamefont {St{\"u}ble}\ \emph {et~al.}(2018)\citenamefont
  {St{\"u}ble}, \citenamefont {Peschke}, \citenamefont {Johrendt},\ and\
  \citenamefont {R{\"o}hr}}]{stuble2018na7}%
  \BibitemOpen
  \bibfield  {author} {\bibinfo {author} {\bibfnamefont {P.}~\bibnamefont
  {St{\"u}ble}}, \bibinfo {author} {\bibfnamefont {S.}~\bibnamefont {Peschke}},
  \bibinfo {author} {\bibfnamefont {D.}~\bibnamefont {Johrendt}}, \ and\
  \bibinfo {author} {\bibfnamefont {C.}~\bibnamefont {R{\"o}hr}},\ }\href
  {https://doi.org/10.1016/j.jssc.2017.10.033} {\bibfield  {journal} {\bibinfo
  {journal} {J. Solid State Chem.}\ }\textbf {\bibinfo {volume} {258}},\
  \bibinfo {pages} {416} (\bibinfo {year} {2018})}\BibitemShut {NoStop}%
\bibitem [{\citenamefont {Ku}\ \emph {et~al.}(2002)\citenamefont {Ku},
  \citenamefont {Rosner}, \citenamefont {Pickett},\ and\ \citenamefont
  {Scalettar}}]{ku2002insulating}%
  \BibitemOpen
  \bibfield  {author} {\bibinfo {author} {\bibfnamefont {W.}~\bibnamefont
  {Ku}}, \bibinfo {author} {\bibfnamefont {H.}~\bibnamefont {Rosner}}, \bibinfo
  {author} {\bibfnamefont {W.~E.}\ \bibnamefont {Pickett}}, \ and\ \bibinfo
  {author} {\bibfnamefont {R.~T.}\ \bibnamefont {Scalettar}},\ }\href
  {https://doi.org/10.1103/PhysRevLett.89.167204} {\bibfield  {journal}
  {\bibinfo  {journal} {Phys. Rev. Lett.}\ }\textbf {\bibinfo {volume} {89}},\
  \bibinfo {pages} {167204} (\bibinfo {year} {2002})}\BibitemShut {NoStop}%
\bibitem [{\citenamefont {Momma}\ and\ \citenamefont
  {Izumi}(2011)}]{momma2011vesta}%
  \BibitemOpen
  \bibfield  {author} {\bibinfo {author} {\bibfnamefont {K.}~\bibnamefont
  {Momma}}\ and\ \bibinfo {author} {\bibfnamefont {F.}~\bibnamefont {Izumi}},\
  }\href {https://doi.org/10.1107/S0021889811038970} {\bibfield  {journal}
  {\bibinfo  {journal} {J. Appl. Crystallogr.}\ }\textbf {\bibinfo {volume}
  {44}},\ \bibinfo {pages} {1272} (\bibinfo {year} {2011})}\BibitemShut
  {NoStop}%
\end{thebibliography}%
\end{document}